\documentclass[a4paper,11pt]{article}
\usepackage{pos}
\usepackage{graphicx}
\usepackage{hyperref}
\usepackage{bm}
\usepackage{amsmath}
\usepackage{amssymb}
\usepackage{natbib}
\usepackage{multicol}

\usepackage{dirtytalk}
\newcommand{\rb}[1]{{\color{black}  #1}} 
\usepackage[normalem]{ulem}
\definecolor{black}{RGB}{0,0,128}

\title{Modeling non-thermal emission from SN 1987A }
 \ShortTitle{Modeling non-thermal emission from SN 1987A }

\author*[a,b]{Robert Brose}
\author[a,c]{Jonathan Mackey}
\author[a,d]{Sean Kelly}
\author[e]{Nathan Grin}
\author[e]{Luca Grassitelli}

\affiliation[a]{Dublin Institute for Advanced Studies\\
  31 Fitzwilliam Place, Dublin 2, Ireland}

\affiliation[b]{Institute of Physics and Astronomy, University of Potsdam\\
14476 Potsdam, Germany}

\affiliation[c]{Centre for AstroParticle Physics and Astrophysics (CAPPA), DIAS Dunsink Observatory\\
Dunsink Lane, Dublin 15, Ireland}

\affiliation[d]{Dublin City University\\
Collins Ave Ext, Whitehall, Dublin 9, Ireland}

\affiliation[e]{Argelander-Institut f\"ur Astronomie, Universität Bonn, Auf dem H\"ugel 71, 53121 Bonn, Germany}

\emailAdd{broserob@cp.dias.ie}
\emailAdd{jmackey@cp.dias.ie}

\abstract{
The remnant of SN 1987A is the best-studied object of its kind. The rich data-set of its thermal and non-thermal emission across the electromagnetic spectrum poses a unique testbed for the elaboration of particle-acceleration theory.

We use 2D simulations of the progenitor's wind to obtain hydro-profiles for the medium around the supernova explosion. Various cones along prominent features of the ambient medium are then used in our time-dependent acceleration code RATPaC to model the evolution of the emission of SN 1987A and compare it to observational data.

We solve for the transport of cosmic rays and the hydrodynamical flow, in the test-particle limit. The simulation code relies on 1D profiles but the large expansion speed of the young remnant renders lateral transport unimportant.

We find that the increase in thermal X-ray emission predates the increase in the low-energy gamma-ray brightness by several years. The increase of the gamma-ray brightness at lower energies is followed by a smooth increase at the highest energies. The gamma-ray spectrum at the highest energies appears soft during the brightening but hardens as more material in the equatorial ring gets shocked. The X-ray and gamma-ray brightness remain almost constant once the SNR blast-wave passed the region of peak-density in the equatorial plane.

}

\FullConference{37$^{\rm{th}}$ International Cosmic Ray Conference (ICRC 2021)\\
		July 12th -- 23rd, 2021\\
		Online -- Berlin, Germany}

\begin{document}
\maketitle

\section{Intoduction}
The remnant of SN 1987A in the large Magellanic cloud is one of the most well observed astronomical objects. It is the first supernova to happen this close to earth since the invention of telescopes, and the only supernova remnant (SNR) that could be observed over its entire lifetime. Further, the progenitor star - a blue supergiant (BSG) called Sanduleak -69$^\circ$ 202 - is known \citep{1987Natur.327..597H}, enabling a connection between the stellar evolution and the circumstellar medium (CSM) shaped by the star and the evolution of the SNR for the first time.

Observations from radio \citep{2002PASA...19..207M}, infrared \citep{2010ApJ...722..425D}, optical \citep{2000ApJ...537L.123L} to X-ray \citep{2011Natur.474..484L, 2013ApJ...764...11H} revealed a rebrightening of the remnant after the initial emission caused by the explosion had faded. This is associated with the interaction of the SNR-ejecta with a dense ring of material in the equatorial plane around the progenitor star. Recently, a brightening at GeV gamma-ray energies was reported as well \citep{2019arXiv190303045M} with a claimed $5\sigma$ detection in the epoch between 2016-2018. However, source confusion is a possible problem given the spatial resolution of Fermi-LAT at this large distance. Further, very young SNRs are considered as sources of the highest energetic cosmic-rays (CR). So far, H.E.S.S. found no sign of very-high energy (VHE) gamma-ray emission from SN 1987A that could be associated with proton acceleration beyond TeV-energies \citep{2015Sci...347..406H}.

There are mainly two mechanisms discussed for the formation of the dense, equatorial ring. One is based on the merger of two massive stars about $20\,000\,$years prior to the explosion \citep{2007Sci...315.1103M}, the other on a spin up to critical rotation of the progenitor star during a transition from a red supergiant (RSG) to a BSG prior to explosion \citep{1998A&A...334..210H}.
In this work we explore the latter scenario, combining 2D simulations of the ambient medium with a one-dimension code for the acceleration of CRs. 

\section{Method}

We first describe our stellar evolution calculations and 2D-simulations of the CSM around the progenitor of SN 1987A.
We use the CSM from these simulations to construct a \say{toy} model for the hydrodynamical structure around SN 1987A.
Afterwards we describe the 1D model employed to study the particle acceleration in the SNR.

\subsection{Stellar evolution and 2D hydrodynamic model}




We used Modules for Experiments in Stellar Astrophysics (MESA) \citep{2011ApJS..192....3P, 2018ApJS..234...34P} to calculate single-star evolutionary sequences for rotating stars with LMC metallicity.
The elemental abundances are taken from \citep{2011A&A...530A.115B}, and the particular evolutionary sequence used is from a star with initial mass $M_i=20\,\mathrm{M}_\odot$ and initial rotation velocity $v_\mathrm{rot}=250$\,km\,s$^{-1}$.
The RSG phase begins at $t\approx9.25$\,Myr and ends when the star embarks on a blue loop at $t\approx9.585$\,Myr. The BSG phase also lasts about 0.35\,Myr, after which the star re-expands to the RSG branch.

\begin{figure}
    \centering
    \begin{minipage}{0.69\textwidth}
    \includegraphics[width=0.99\textwidth]{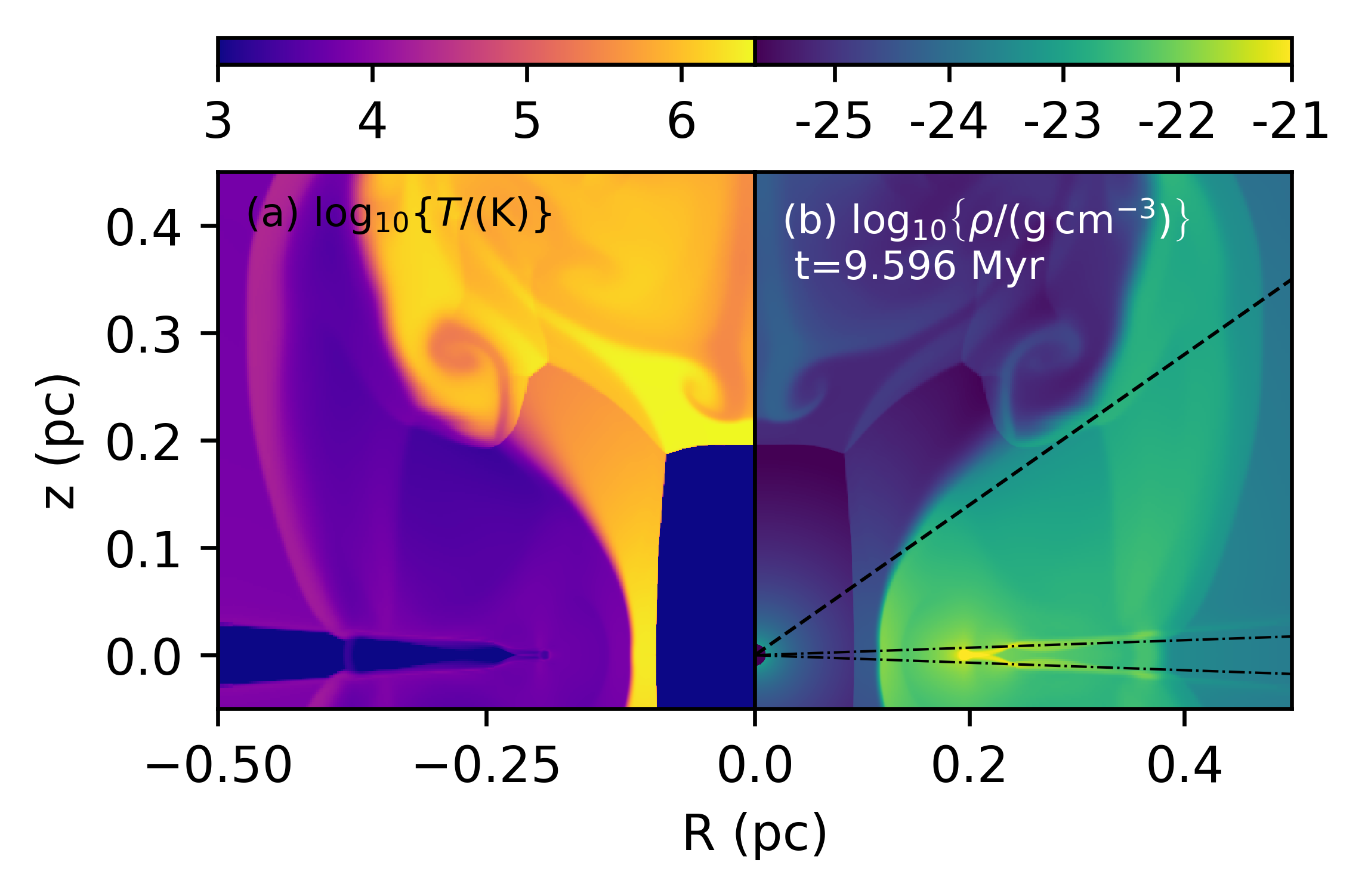}
    \end{minipage}
    \begin{minipage}{0.29\textwidth}
    \caption{Gas temperature (a) and density (b) showing the equatorial region and upper half-plane of a 2D axisymmetric simulation of wind-wind interaction as a star evolves from RSG to BSG, spinning up to critical rotation in the process.
    The CSM is shown 10\,000\,yr after the critical rotation phase.
    Dot-dashed lines are $\pm2^\circ$, and the dashed line 35$^\circ$, from the equatorial plane.}
    \label{fig:Pion}
    \end{minipage}
\end{figure}

This evolutionary sequence might not be representative of the progenitor of SN 1987A because the blue loop begins almost 0.4\,Myr before the end of the stars life.
The nebula associated with the red-to-blue stellar transition would be long gone after such a long time, driven away by the fast wind of the BSG phase.
The progenitor of SN 1987A, in contrast, exploded as a BSG, and the kinematic age of the circumstellar rings is $\sim 10^4$\,years, which should correspond to the length of time between the critical rotation phase and core collapse.
The timing of the red-to-blue transition is extremely sensitive to many of the tunable parameters in stellar evolution codes, and so the fact that this calculation embarked on a blue loop 0.35\,Myr  (and not \rb{0.025 Myr}) before the end of its life is not a strong prediction of any stellar evolution code. 
There are calculations where the star goes on a blue loop just before the end of its life \citep{1991A&A...252..669L}. Nevertheless, this evolutionary sequence is instructive in determining whether such a red-to-blue transition for a rotating star could, in principle, produce a nebula similar to that seen in SN987A.

We modified the mass-loss history of the calculation in two ways: reducing the mass-loss rate after critical rotation, and increasing it around the critical rotation phase.
After critical rotation we reduced the mass-loss rate linearly from the standard rate to a rate $10\times$ lower over a period of $10^{10}$\,s.
This was to reduce the ram pressure of the wind, thereby allowing a larger H~\textsc{ii} region to develop around the equatorial ring (see below and \citep{1995ApJ...452L..45C}).
Such low rates have some observational support in the weak wind measured for the apparent Galactic twin of the SN 1987A progenitor, SBW1 \citep{2017MNRAS.468.2333S}.
During the phase of near-critical rotation, when $\Omega\equiv v_\mathrm{rot}/v_\mathrm{crit} > 0.5$, we linearly increased $\dot{M}$ according to the following presciption:
\begin{equation}
    \dot{M}_1 = \dot{M}_0 \left(1 + 4\frac{\Omega-0.5}{0.5}\right) \;
\end{equation}
where $\dot{M}_1$ is the enhanced mass-loss rate with respect to the MESA rate, $\dot{M}_0$ (using the ``Dutch'' mass-loss scheme \rb{\citep{1988A&AS...72..259D,2001A&A...369..574V}}).
This prescription increases $\dot{M}$ by a factor of 5 at critical rotation; in the MESA calculation, the total mass lost is thought to be regulated by the angular momentum loss needed for the contracting star to remain at sub-critical rotation.
Nevertheless, the details of mass loss at critical rotation are poorly understood and the observed mass in the equatorial ring of SN 1987A is evidence that significant mass can be lost.
The ad-hoc modification ensures that sufficient mass is in the equatorial ring to agree with observations.

We performed 2D radiation-magnetohydrodynamic (MHD) simulations using \textsc{PION} \citep{Mackey2021} with the above-mentioned stellar evolution model providing an inner boundary condition, to generate a circumstellar nebula similar to that of SN 1987A. 
Focusing of the wind towards the equator at phases of near-critical rotation follows \citep{1999ApJ...520L..49L, Mackey2021}.
Figure \ref{fig:Pion} shows a dense equatorial ring $\approx0.2\,$pc from the central star, surrounded by a photoevaporation flow expanding from the ring.
This flow produces a photoionized H~\textsc{ii} region between the BSG wind and the equatatorial ring \citep{1995ApJ...452L..45C}.
The structure is similar to the model proposed by \citet{2012ApJ...752..103D}, with the difference that the inner ring has a somewhat larger extension, $\approx\pm2^\circ$ vs. $\pm0.1^\circ$, as well as the outer ring $\approx\pm30^\circ$ vs. $\pm15^\circ$.

\subsection{Toy model}
\begin{figure}
    \centering
    \begin{minipage}{0.69\textwidth}
    \includegraphics[width=0.99\textwidth]{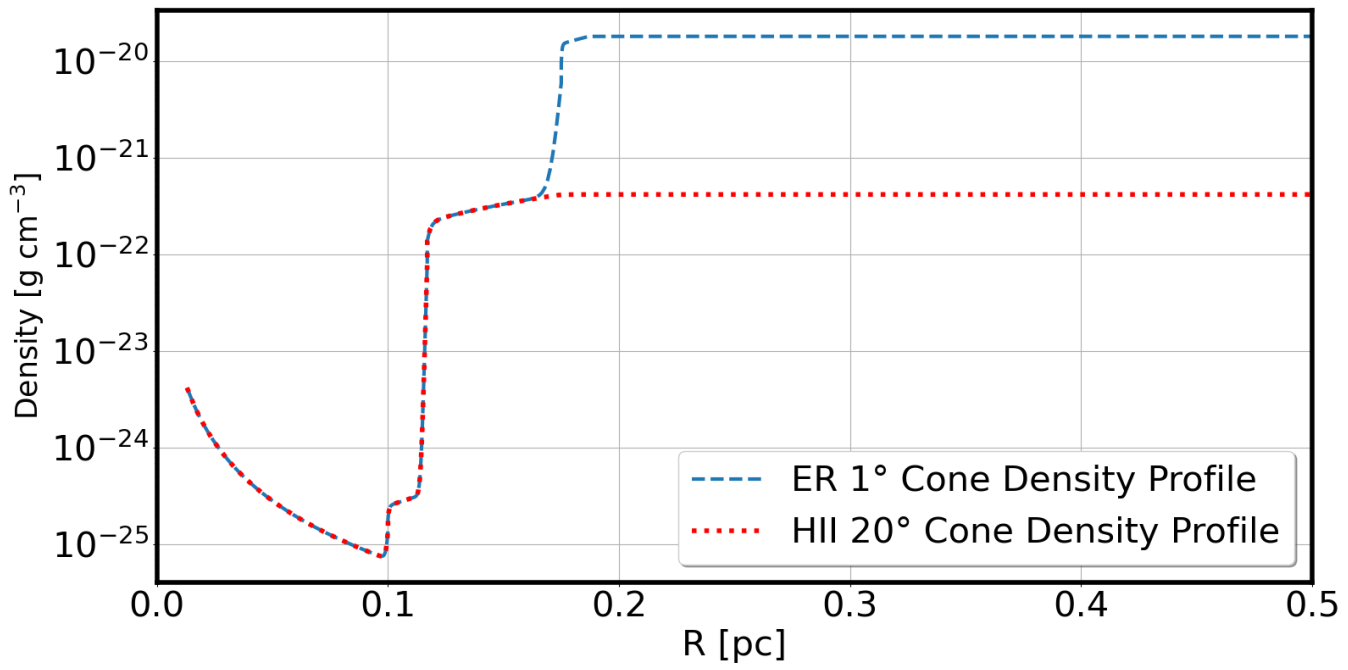}
    \end{minipage}
    \begin{minipage}{0.29\textwidth}    
    \caption{Density structure of the two cones used for the modeling of the soft X-ray emission and the acceleration of CRs. }
    \label{fig:density}
    \end{minipage}
\end{figure}

Based on the results from the previous section, we constructed a \say{toy}-model (Figure \ref{fig:density}) that features an outer cone with an extension of $\pm20^\circ$ and an inner cone with a high-density equatorial ring with a cone-width of $\pm1^\circ$. 
We \rb{set} the densities - $3\times10^{-22}\,$g/cm$^{-3}$ in the H~\textsc{ii} region and $2\times10^{-20}\,$g/cm$^{-3}$ in the dense ring - in order to reproduce the observed soft X-ray lightcurve of SN 1987A (see also Figure \ref{fig:fluxes}).
In this \say{toy}-model we considered a constant density in the dense equatorial ring. The density distribution derived from the MHD-simulation suggests that the density will decrease after reaching the peak, which is also more in line with the evolution of the thermal X-ray flux in recent years \citep{2016ApJ...829...40F}.

The X-ray emission is calculated from the gas density and temperature using the approximations of Hnatyk \& Petruk \citep{1999A&A...344..295H}. We note that this calculation of the continuum thermal X-ray emission assuming local thermal equilibrium is not strictly valid in the case of SN 1987A given the additional heating and ionization shortly after the supernova explosion by the radiation fields of the progenitor star. However, a detailed modeling of the X-ray emission is beyond the scope of this paper and the derived emission is  a crude estimate to tailor the density-profiles used for the CR-acceleration code.

\subsection{Particle acceleration}
We use the \textbf{R}adiation \textbf{A}cceleration \textbf{T}ransport \textbf{Pa}rallel \textbf{C}ode (RATPaC) to calculate the particle acceleration and subsequent thermal and non-thermal emission. A detailed description of the code can be found here: \cite{2012APh....35..300T, 2013A&A...552A.102T, 2018A&A...618A.155S, 2019A&A...627A.166B}. 

\subsubsection{Cosmic-rays}
We solve the kinetic equation for the transport of CRs in the form 
\begin{equation}
\frac{\partial N}{\partial t}=\nabla(D\nabla N-\vec{v}N)-\frac{\partial}{\partial p}\left((N\dot{p})-\frac{\nabla \vec{v}}{3}Np\right)+Q,
\label{Transport}
\end{equation}
where $N$ is the differential number density of cosmic rays, $D$ is the spatial diffusion coefficient, $\vec{v}$ is the plasma velocity, $\dot p$ represents the energy losses (in our case synchrotron losses) and $Q$ is the source term.


We assume the thermal leakage injection model \citep{2005MNRAS.361..907B}, where $p_{\rm inj}$ is the minimum momentum for which a thermal particle can cross the shock and enter the acceleration process and the injection efficiency for the compression ratio of 4 is determined as
\begin{equation}
  \eta = \frac{4}{\sqrt{\pi}}\frac{\xi^3}{e^{\xi^2}}.
\end{equation}

We solve the transport equation for protons in 1D using the RATPaC code as described in \cite{2012APh....35..300T, 2013A&A...552A.102T, 2018A&A...618A.155S, 2019A&A...627A.166B} taking into account only a forward shock and ignoring the reverse shock. Resulting proton spectra at different epochs are then used to calculate the Pion-decay emission from the remnant. In this study we do not consider acceleration of electrons and also assume that cosmic-rays will not dynamically effect the SNR evolution.

We apply Bohm-like diffusion in the interior of the remnant assuming a simple magnetic-field geometry with a constant field in the upstream and downstream respectively. The field in the upstream is assumed to be/amplified to $125\,\mu$G and the field in the downstream has a constant value of $410\,\mu$G. 


\subsubsection{Hydrodynamic evolution}
The standard gasdynamical equations
\begin{align}
\frac{\partial }{\partial t}\left( \begin{array}{c}
                                    \rho\\
				    \vec{m}\\
				    E
                               \end{array}
 \right) + \nabla\left( \begin{array}{c}
                   \rho\vec{v}\\
		   \vec{mv} + P\vec{I}\\
		   (E+P)\vec{v} 
                 \end{array}
 \right)^T = \left(\begin{array}{c}
            0\\
		    0\\
		    0
            \end{array}\right),
 \,\,\,\,\,\,\,\,\,\,\,\frac{\rho\vec{v}^2}{2}+\frac{P}{\gamma-1}  = E 
\end{align}
are solved, where $\rho$ is the density of the thermal gas, $\vec{v}$ the plasma velocity, $\vec{m}=\vec{v}\rho$ the momentum density, $P$ the thermal pressure of the gas, $\vec{I}$ the unit tensor, and $E$ the total energy of the ideal gas with $\gamma=5/3$. We assume that the magnetic field is dynamically unimportant due to its low strength and the remnant not being in the radiative phase yet \citep{2016MNRAS.456.2343P}. The equations are solved in 1D for a spherical symmetry.

We initialize the ejecta profile by a plateau in density with the value $\rho_{\mathrm c}$ up to the radius $r_{\mathrm c}$ followed by a power-law distribution up to the ejecta-radius $R_{\mathrm{ej}}$:

\begin{align}
 \rho \left(r\right) &= \begin{cases}
             \rho_{\mathrm c}, & r<r_{\mathrm c},\\
             \rho_{\mathrm c}\left(\frac{r}{r_{\mathrm c}}\right)^{-n}, & r_{\mathrm c}\leq r \leq R_{\mathrm{ej}},\\
             \rho_\text{CSM}\left(r\right), & r>R_{\mathrm{ej}}.\\
            \end{cases}
	\label{gasdyn}
\end{align}

The exponent for the ejecta profile is set to $n=8$ for SN 1987A \citep{2015ApJ...810..168O}. The velocity of the ejecta is defined as
\begin{equation}
v_{\mathrm{ej}}\left(r\right) = \frac{r}{T_{\mathrm{SN}}},
\end{equation}
where $T_{\mathrm{SN}}=0.5$\,yr is the initial time set for hydrodynamic simulations. We assumed an ejecta-mass of $M_{\mathrm{ej}}=14\,M_\odot$ and an explosion-energy of $E_{\mathrm{ej}}=1.5\times10^{51}\,$erg \cite{2007AIPC..937...25U}.
The initial temperature is set to $10^4$~K everywhere and the initial pressure is calculated using the equation of state. We used $130,000$ linearly distributed grid-points and $2\,$pc grid-size for the hydro simulations. 

We further assume in Eq.~\ref{gasdyn} that the SNR is expanding into the CSM created by the progenitor star. We read in the density, velocity and pressure distribution from our \say{toy}-model. There are two cones considered, the outer cone featuring the H~\textsc{ii} region only and the inner cone also containing the material in the dense equatorial ring. The simulation is run independently for both cones and then the emission from both cones is combined accounting for the different fractional surface areas associated with the opening angles.  



%

\section{Results}
The \say{toy}-model that we constructed based on our 2D MHD simulations reasonably well reproduces the observed soft X-ray light curve (Figure \ref{fig:fluxes}). However, here are two things to note. The first is that our simulations can not account for the suppressed hard X-ray flux as soon as the SNR-shock starts interacting with the dense equatorial ring. The main reason is, that therefore additional clumping needs to be assumed in order to generate a plasma-component with a low-enough temperature \citep{2012ApJ...752..103D}. These clumps could further enhance the gamma-ray luminosity of SN 1987A by providing additional target material for p-p-interactions.
Secondly, we assumed the equatorial ring to stay dense. However, a reduction of the density is observed in the MHD-simulations and supported by the flattening of the X-ray light-curve of SN 1987A \citep{2016ApJ...829...40F}. 

\subsection{Gamma-ray flux}
\begin{figure}
    \centering
    \begin{minipage}{0.69\textwidth}
    \includegraphics[width=0.9\textwidth]{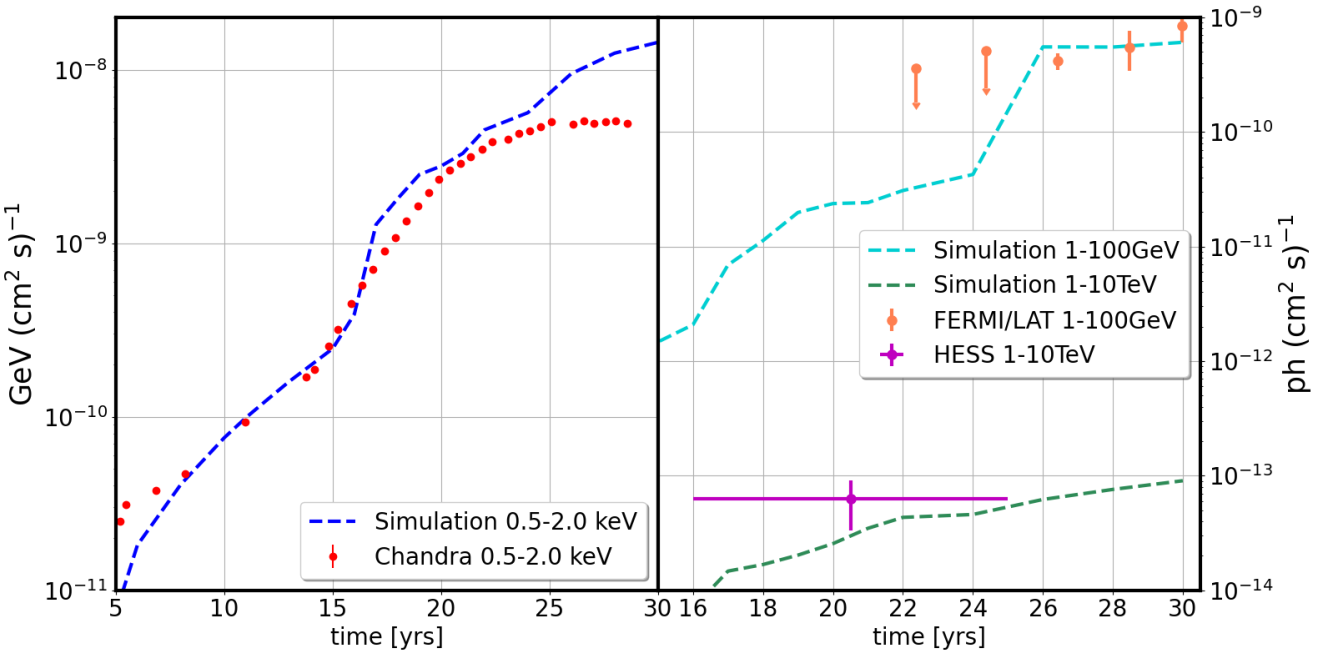}
    \end{minipage}
    \begin{minipage}{0.29\textwidth}
    \caption{Left hand side: Thermal X-ray flux of SN 1987A based on the emission from the two cones, weighted by their surface area, compared to Chandra observations \citep{2016ApJ...829...40F}. Right hand side: Photon Fluxes in the Fermi-LAT \citep{2019arXiv190303045M} and H.E.S.S. \citep{2015Sci...347..406H} energy ranges.}
    \label{fig:fluxes}
    \end{minipage}
\end{figure}

Figure \ref{fig:fluxes} shows the gamma-ray photon fluxes in the  Fermi-LAT high-energy (1-100GeV)  and H.E.S.S. very-high energy (1-10TeV) ranges compared to the predictions from our model.
It is evident that the high-energy flux increases by a factor of $\approx100$ between 2002 and 2017. There is a sharp increase by a factor of $\approx10$ in 2012, when the interaction with the dense equatorial ring translates into an increased gamma-ray flux. There is a delay of $\approx8\,$years between the accelerated brightening in X-rays and the brightening in high-energy gamma-rays. The reason is that the low-energy CRs responsible for the emission in the Fermi-LAT domain are advectively transported away from the shock with the flow. They are not able to return easily when the shock interacts with the dense material. Thus CRs need to be freshly accelerated beyond $100\,$GeV to produce gamma-rays detectable by Fermi-LAT. The shock-speed is reduced to the order of $1000\,$km\,s$^{-1}$ by the interaction with the dense material, greatly increasing the acceleration time of CRs.

This increased acceleration time is also responsible for the lack of a strong increase of the very-high energy gamma-ray luminosity. The reduced acceleration efficiency prevents the acceleration of freshly injected CRs to relevant energies (see also Figure \ref{fig:spec}). However, the gamma-ray flux in the H.E.S.S.-domain is expected to increase by a factor of $\approx10$ between 2002 and 2017 on account of the expansion of the remnant. Further, CRs with energies beyond a TeV can more easily diffuse back to the shock and probe the higher density there, once the SNR started to interact with the dense equatorial ring. 

\subsection{Gamma-ray spectra}
\begin{figure}
    \centering
    \begin{minipage}{0.69\textwidth}
    \includegraphics[width=0.9\textwidth]{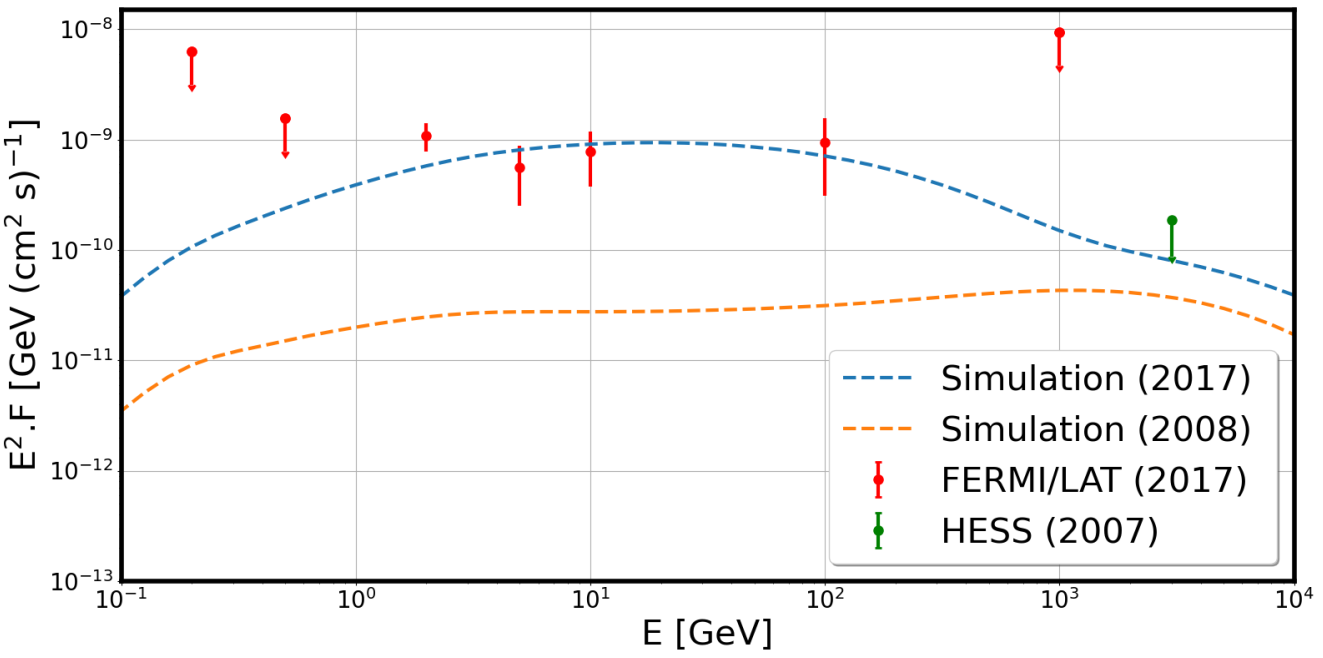}
    \end{minipage}
    \begin{minipage}{0.29\textwidth}
    \caption{Comparison of the simulated gamma-ray spectra between 2008 and 2017 and Fermi-LAT \citep{2019arXiv190303045M} and H.E.S.S.\ observations \citep{2015Sci...347..406H}.}
    \label{fig:spec}
    \end{minipage}
\end{figure}

The mechanisms explained in the previous section are also evident in the comparison of the gamma-ray spectra between 2008 and 2017 shown in Figure \ref{fig:spec}.
It is evident that none of the particles injected after the interaction with the dense equatorial ring have been accelerated to energies relevant for H.E.S.S. by 2017. However, the shock propagating in the outer cone, associated with the H~\textsc{ii} region, is increasing in size and still moving at a high velocity, so that acceleration to post-TeV energies is taking place. Thus, the increase in brightness at very-high energies is still solely powered by this intermediate density region.

The transition from the freshly accelerated low-energy CRs from the dense equatorial ring and the high-energy CRs from the H~\textsc{ii} region provide a soft spectrum at energies beyond a few hundreds of GeV with a spectral index of $s\approx2.6$ at these energies. That is considerably softer than predicted by standard DSA. 

It has to be noted that the injection efficiency that we assumed for our models is the value that is roughly consistent with injection fractions observed in historical remnants. Formally, the injection fraction is a free parameter in acceleration models and, in case of the thermal-leakage model, is associated with the multiple of the thermal momentum that particles need to have in order to get accelerated as CRs by the shock. However, we like to note that the additional, clumpy material that is needed in the equatorial ring in order to explain the hard X-ray emission will also enhance the gamma-ray luminosity from that region. A similar model has been proposed by Gabici \& Aharonian \citep{2014MNRAS.445L..70G} for RX J1713.7-3946. However, there is a lack of intermediate-temperature X-ray emission from the shock-crushed dense clouds in case of RX J1713.7-3946. For SN 1987A, material at corresponding temperatures has been detected. Further, the observed optical knots in the equatorial ring of SN 1987A provide already experimental evidence for a significant amount of clumping, which might need to be taken into account additionally when modeling the gamma-ray emission.

\section{Conclusions}
We used \textsc{PION} to perform 2D MHD simulations based on a red-to-blue transition of the rotating progenitor star of SN 1987A to derive a model for the CSM. We constructed a toy-model based on these MHD simulations, with two cones containing (i) an intermediate density H~\textsc{ii} region and (ii) an intermediate density H~\textsc{ii} region plus a dense equatorial ring.
Using this model we calculated the particle acceleration and gamma-ray emission from SN 1987A using RATPaC.
\begin{itemize}
    \item We find a good agreement between our derived density-distribution and observed soft X-ray emission.
    \item Our model predicts a significant brightening at high-energy gamma-rays between 2002 and 2017 with a strong increase around 2012. The predicted fluxes are consistent with the recent marginal detection of SN 1987A in Fermi-LAT data.
    \item We predict an increase of the very-high energy gamma-ray luminosity by a factor of $\approx3$ between 2002 and 2017 powered by the acceleration in the intermediate-density H~\textsc{ii} region.
\end{itemize}

\begin{multicols}{4}
\scriptsize
\bibliographystyle{aa}
\bibliography{References.bib}
\end{multicols}

\end{document}